\documentclass[useAMS,usenatbib]{mn2e}
\usepackage{epsfig}
\usepackage[totalwidth=480pt, totalheight=680pt]{geometry} 

\def\beq{\begin{equation}}
\def\eeq{\end{equation}}

\def\erf{\mathrm{erf}}
\def\media#1{\langle #1\rangle}

\def\nod{...}
\def\msol{\,\mathrm{M}_\odot}
\def\M{M_\bullet}

\def\Mmin{M_{\bullet,1}}
\def\Mmax{M_{\bullet,2}}
\def\wbh{W_\mathrm{BH}}

\def\rbh{r_\mathrm{BH}}
\def\xbh{x_\mathrm{BH}}
\def\e{\mathrm{e}}
\def\ud{\mathrm{d}}
\def\textchanged#1{#1}

\title[Intermediate-mass black holes in globular clusters and EHB stars]
{The presence of intermediate-mass black holes in globular clusters and
their connection with extreme horizontal branch stars}
\author[P. Miocchi]{P. Miocchi\thanks{E-mail:
miocchi@uniroma1.it}\\
INAF, Osservatorio Astronomico di Teramo,
via M. Maggini, Teramo I-64100, Italy.\\
Dipartimento di Fisica, Universit\'a di Roma ``La Sapienza'', P.le A. Moro, 5, Roma I-00185, Italy.}
\begin{document}

\date{Accepted ????. Received ????; in original form ????}

\pagerange{\pageref{firstpage}--\pageref{lastpage}} \pubyear{????}

\maketitle

\label{firstpage}

\begin{abstract}
By means of a multimass isotropic and spherical model including
self-consistently a central intermediate-mass black hole (IMBH),
the influence of this object on the morphological and physical properties of
globular clusters is investigated in this paper.
Confirming recent numerical studies, it is found that a cluster (with mass $M$)
hosting an IMBH (with mass $\M$) shows, outside the black hole gravitational
influence region, a core-like profile resembling a King profile with concentration
$c<2$, though with a slightly steeper behaviour in the core region.
In particular, the core logarithmic slope is $s\la 0.25$ for reasonably low IMBH masses
($\M\la 10^{-2}M$), while $c$ decreases monotonically with $\M$.
Completely power-law density profiles (similar to, e.g.,
that of collapsed clusters) are admitted only in the presence of a
black hole with an unrealistic $\M>M$.
The mass range estimate $12s-4.8<\log (\M/M)<-1.1c - 0.69$, depending on the
morphological parameters, is deduced
considering a wide grid of models.
Applying this estimate to a set of 39 globular clusters,
\textchanged{
it is found that NGC 2808, NGC 6388, M80, M13, M62, M54 and G1 (in M31)
}
probably host an IMBH.
For them, the scaling laws $\M\sim 0.09(M/\msol)^{0.7} \msol$ and
$\M\sim 50(\sigma_\mathrm{obs}/\textrm{km s}^{-1})^{1.2}
\msol$, are identified from weighted least-squares fit.
An important result of this `collective' study is that a strong
correlation exists between the presence of an extreme blue horizontal branch (HB)
and the presence of an IMBH.
\textchanged{
In particular, the presence of a central IMBH in M13 and NGC 6388 could explain
why these clusters possess extreme HB stars, contrarily to the `second parameter
counterparts' M3 and 47 Tuc.
}
\end{abstract}

\defcitealias{noyola06}{NG06}
\defcitealias{BW}{BW76}

\section{Introduction}
The presence and origin of intermediate-mass black holes (IMBHs)
at the centre of globular clusters, are still being debated
inside the astrophysical community \citep[see the recent general reviews in][]
{vandermarel,praga}.
There are essentially two main formation theories:
they could be Population III stellar remnants \citep[see, e.g.,][]{madau} or
could form in a runaway merging of young stars
in sufficiently dense clusters (e.g., \citealt{porteg04}; \citealt*{gurkan};
\citealt*{freitag064}).

The IMBHs existence is suggested primarily by an argument of plausibility:
they would fill the `embarrassing' gap between
stellar black holes (BHs; with mass $\M\la 10\msol$) and super-massive BHs
\textchanged{
(having $\M\sim 10^5$--$10^9 \msol$) 
}
that are believed to reside in the nucleus of most galaxies
\citep[e.g.,][for recent reviews]{kormendy,richstone,ferrarese05}.
By exploiting the accurate kinematical measurements of gas and stars in nearby galaxies,
important correlations between the mass of super-massive BHs and various global
properties of their hosting galaxy have been deduced
\textchanged{
\citep[see][and references therein]{ferrarese05}.
}
For instance, extrapolating the \citet{magorrian} relation -- $\M \sim 10^{-3}$ times
the mass of the host system -- to globular clusters, these would contain BHs with an
intermediate mass $\M\sim 10^2$--$10^4$.
Indirect observational evidences come from the
detection of ultra-luminous X-ray sources radiating at super-Eddington luminosity
($> 10^{39}$ erg s$^{-1}$) and thought to originate from matter infall on BHs
considerably more massive than stellar BHs \citep[see][for a review on this subject]
{fabbiano}.

In globular clusters, however, direct kinematic observations are seriously obstacled
by crowding and by the relatively small number of stars inside the BH
gravitational influence
region (hereafter BHIR), which cannot be well resolved in many cases.
Thus, so far, only very few clusters have been the object of accurate kinematic studies so
as to infer about the possible presence of massive objects in the central regions.

We just remind the cases of the core-collapsed clusters M15 (\citealt{gerssen02,gerssen03};
\citealt*{mcnamara}) and
47 Tuc, in which such a presence is not confirmed yet by the latest
observational data (\citealt{vandenbosch} and
\citealt{mclaughlin}, respectively), and the highly concentrated and massive cluster
G1 (in M31),
where a central $\sim 2\times 10^4 \msol$ IMBH has been claimed to reside
\citep*{geb02, geb05}, while a recent deep photometric analysis of the core
region of $\omega$ Cen seems to be consistent with an IMBH with mass $\sim 10^4 \msol$
\citep*{noyola07}.
Note, however, that for both M15 and G1, direct and accurate $N$-body
simulations reproduce the observed mass-to-light ratio without the need of an IMBH
\citep{bau03a,bau03b},
\textchanged{
although the peculiar scenario of
a two clusters merging event is required in the case of G1.
}

Other, less direct, insights can be gained from the time behaviour of the period
of millisecond pulsars sited in clusters central region, which permits
to deduce the local acceleration. This occurs in the intriguing case
of NGC 6752 where there are indications of the existence of an underluminous
and compact component with mass $\sim 10^3 \msol$
\citep[][and references therein]{ferraro03,colpi05}.

It is interesting to note that the above-mentioned individual studies concerned
globular clusters that are supposed to host IMBH because of their
(present) high central density and (presently) high rate of stellar collisions
in their compact cores.
Nevertheless, present conditions may
be drastically different from those at the early epochs of clusters life.
More importantly, as firstly argued by \citet*{bau05}, their dynamical status
could be even inconsistent with the long-term effects that a IMBH actually
produces on the cluster internal evolution. Using collisional
$N$-body simulations, \citeauthor{bau05} found that the high stellar
density in the vicinity of the central IMBH enhances the rate of
close encounters that, in turn, induces a rapid expansion of the central region,
giving rise to a medium-concentration, King-like profile, whose features are
almost independent of the BH mass. They argued that an observable
fingerprint of the IMBH presence is just a slight slope of the density in the core region,
thus suggesting as probable candidate clusters hosting IMBH, those having a surface
brightness logarithmic slope $\sim -0.2$ in the core (while the typical
projected profile of core-collapsed systems goes like $r^{-1}$).

Similar conclusions have been recently reached by \citet{trenti07} who
conducted direct $N$-body experiments of clusters with IMBHs and
realistic fractions of primordial binaries: the expansion of the core is further
enhanced and the
formation of the density cusp is confined in the immediate vicinity of the BH.
This leads to density profiles with a core to half-mass radius ratio
significantly higher than when the IMBH is absent \citep{trenti06b}.
All these studies are telling us that a IMBH represents a `heat' source
in the central region, acting in a way similar to hardening binaries in
contrasting gravothermal collapse \citep[see also][]{heggiehut03}.

However, besides these global morphological hints, other fingerprints
of the IMBH presence could be revealed. As known, one of the most important effect
of the gravitational influence of the IMBH on the surrounding stars, is the tidal
erosion and disruption they undergo
during close passages (see, e.g., \citealt{frank76}; \citealt*{pau}; \citealt*{bau04a};
\citealt*{freitag063}; \citealt{bau06}).
Besides complete disaggregation, this may lead to a loss of envelope mass of
passing-by giants. The mass loss from stellar outer layers is also one of the possible
explanation for the not yet ascertained origin of `blue subdwarfs' -- also called
extreme horizontal branch stars, see \citealt{rich97} and, for a review
on this subject in the context of dense galactic nuclei, also
\citealt{alex05}, section 3.4.
These are stars heavier and bluer than turn-off stars, and they are located at the lower-left
end of the horizontal branch (HB), in the region of the color-magnitude diagram
referred to as extreme HB (hereafter EHB; see the discussion in
\citealt{heggiemeylan}, sect. 9.7, and the recent review
in \citealt{catelan05}).

Given the cuspy behaviour of the stellar density in the BHIR and the
relatively high ratio between $\M$ and the single stellar mass, it is
reasonable to expect that the above-mentioned mechanism of blue subdwarfs formation is significantly
enhanced by the presence of the IMBH. Thus, the possibility to find some connection
between this presence and that of an abundant population of EHB stars
naturally emerges and, to our knowledge, has never been explored so far.
This is one of the aim of the present paper.

While the direct investigation of the dynamical evolution of a cluster harbouring
a massive compact object is still at the beginning -- because of the recent development
of sufficiently advanced computational tools -- the study of the stellar phase-space
distribution function (DF) around a massive BH in the quasi-steady regime is a classical
subject in Stellar Dynamics tackled by several authors since the early 70s
\citep{peebles72,BW,sha76,light77,cohn78}. If a time much longer than the
cluster relaxation time is passed since the BH formation and accretion,
dynamical and `thermal' equilibrium can take place and a Maxwellian
DF $\sim \exp(-E/\sigma^2)$, with $E$ the total energy and $\sigma$ a velocity parameter,
should be established. Nevertheless, close
enough to the compact object, stars are either disrupted, because of the strong
tidal interaction, or swallowed by the BH, thus generating an outward flux of
positive energy that, in stationary conditions, must be constant and uniform
through the central region.
This precludes the Maxwellian distribution being valid in the BH vicinity.
Simple scaling arguments \citep[][section 8.4.7]{BT} or
treatments based on the `heat-transfer' equation \citep[][section 2.2]{freitag064} show, indeed,
that a realistic space density behaviour is $\sim r^{-7/4}$.

Such a power-law density (or `cusp') corresponds to a DF $\sim (-E)^{1/4}$,
as was first found, through approximate solutions of the Fokker-Planck equation,
by \citet{BW} \citepalias[hereafter][]{BW} and more recently confirmed with the help of more
accurate methods including relaxation effects, such as: Monte-Carlo codes \citep{sha85, freitag02},
gas-dynamical methods \citep{pau} and direct $N$-body simulations
(\citealt{bau04a}; \citealt*{preto}).

However, equilibrium self-consistent models of
stellar systems including massive BHs are not so numerous.
We just mention the \citet{gerssen02} paper on M15 employing spherical
non-parametric models and the papers by \citet{vandenbosch} and \citet{vandeven},
in which axisymmetric models of, respectively, M15 and $\omega$ Cen are
generated to infer the radial behaviour of the $M/L$ ratio,
by means of a generalization of the \citet{schwa}
orbit superposition method. A further generalization of this technique has been
also recently presented in \citet{alex} to model triaxial galaxies with dark matter
and a central density cusp. It is worth noting, finally, that
\citet{mclaughlin}, in order to derive hints about the
presence of an IMBH in 47 Tuc, applied a single-mass and isotropic King model,
even if in a not fully self-consistent way.

In the present study, a self-consistent parametric model is constructed and used to estimate
the mass of possible IMBHs in clusters and their influence on the cluster density
profile, as well as on the presence of EHB stars. It basically consists of a
multimass King model appropriately extended to incorporate the \citetalias{BW}
DF inside the BHIR.
The model is spherical, isotropic (in the present version) and relatively easy to
construct and use.
Its multimass nature makes straightforward the inclusion of central energy equipartition,
which is suggested to play a non-negligible role in the presence of a massive BH, too
(\citealt{BW77}; \citealt*{murphy91,bau04b}; \citealt{freitag064}; \citealt{hopman06a}).

The model will be described in section \ref{descr}, while in section \ref{profiles} the observable
features of the generated surface brightness and velocity dispersion profiles will be
illustrated, also with respect to the debated structural properties of clusters
hosting IMBH. Then, in section \ref{massest}, the model will be employed to
deduce the mass of possible IMBHs from the central density profiles
provided by the recent \citet{noyola06}
\citepalias[hereafter][]{noyola06} analysis of WFPC2 \textit{Hubble Space Telescope}
(\textit{HST}) observations of 38 galactic globulars (G1 in M31 is considered
as well).
In section \ref{ebtconnection}, finally,
the connection between the existence of IMBHs and
the presence of EHBs will be investigated. Conclusions will be drawn in section \ref{concl}.

\section{Description of the parametric model}
\label{descr}
In the following, given the length-, velocity- and density-scale
parameters, $\tilde r$, $\sigma$ and $\tilde \rho$,
respectively, it is convenient to deal with dimensionless quantities,
namely:
\begin{description}
\item radius $x\equiv r/\tilde r$;
\item velocity $w\equiv v/\sigma$;
\item mean {\it total} potential $W(x)\equiv -\Psi(x)/\sigma^2$;
\item energy (per unit mass) $E\equiv w^2/2-W$;
\item density $\nu(x)\equiv\rho(x)/\tilde\rho$;
\item cluster mass in stars inside the radius $x$, ${\cal M}(x)\equiv M(x)/\tilde \rho\tilde r^3$.
\item BH mass $\mu\equiv \M/\tilde \rho\tilde r^3$.
\end{description}

A suitable DF can be given by generalizing
the \citet{king66} lowered Maxwellian, so as to include
the \citetalias{BW} DF below a proper `transition' energy, i.e.
\beq
f(E)=\left\{
\begin{array}{ll}
 c (-E)^{1/4},&
\textrm{if $E<-\wbh$,}\\
 (2\pi)^{-3/2}(\e^{-E}-1),&
\textrm{if $-\wbh\le E<0$,}\\
 0, & \textrm{if $E\geq 0$,}
\end{array}\right.
\label{DF}
\eeq
where the coefficient $c\equiv (2\pi)^{-3/2}(\e^{\wbh} -1)\wbh^{-1/4}$
assures the $f$ continuity and $W$ is defined so to have $W(x_\mathrm{t})=0$
where $x_\mathrm{t}$ is the model `limiting radius'.
The `transition' potential $\wbh\equiv W(\xbh)$ is considered to be the potential
on the surface of the sphere, with radius $\xbh$, coinciding with the BHIR.

The BHIR can be defined as the largest sphere,
centered at the BH position, in which the motion of
the stars is dominated by the field generated by the compact object.
This definition is coherent with the assumption of Maxwellian behaviour
just outside this region (i.e., for $E\ge -\wbh$).
Then, a reasonable condition for $\xbh$ is that the enclosed mass
in stars is a small fraction of the BH mass, i.e.
\beq
{\cal M}(\xbh) =0.1\mu, \label{xbh0}
\eeq
even if this forces to solve (as we will see later) an implicit equation that
involves a mass profile that is only a-posteriori known.
This condition is similar to that adopted in \citet{merritt}; nevertheless
we chose a smaller fraction ($0.1$ instead of $2$) to ensure the validity
of the power-law behaviour for $f(E)$ \citepalias[see][section IIIc]{BW}.

The usually adopted formula $\xbh\sim G\M/\langle v^2\rangle\tilde r$
(\citetalias{BW}; \citealt{bau04a,bau04b})
is not a-priori compatible with this BHIR definition, because it gives a
radius that yields typically ${\cal M}(\xbh) \la 10^{-4}\mu$.
Indeed, even if the BHIR has to be sufficiently small for the BH gravitational
field to dominate the dynamics inside, on the other hand it must be large enough
to make plausible the hypothesis of isothermal distribution outside it.
This latter condition would not be realistic in an environment in which the BH still
exerts a strong influence on the dynamics \citepalias{BW}.

The DF of equation~(\ref{DF}) guarantees the `dynamical equilibrium' of the system, in the
sense that, according to the Jeans theorem, it is a solution
of the collisionless Boltzmann equation.
The presence of a massive BH, assumed to be located {\it at rest\/}
in the cluster centre, makes that the potential time-independence is
preserved and thus the DF of equation~(\ref{DF}) is still a valid
equilibrium solution, because $E$ remains an integral of motion.
Therefore, one can proceed in the same manner as in obtaining usual
King models, that is -- as required by the self-consistency -- just
by solving, for $W(x)$, the dimensionless Poisson's equation in
spherical symmetry
\beq
\frac{d^2 W}{dx^2}+\frac{2}{x}\frac{dW}{dx}=-(4\pi G\tilde\rho\tilde r^2\sigma^{-2})
\nu(W),
\textrm{ for }x>0,
\label{poisson0}
\eeq
where the origin
is excluded to avoid the BH singularity and
the stellar density, expressed as a function of $W$, is
\beq
\nu(W)=
\ \ 4\pi\int_0^{\sqrt{2W}}f(E)w^2\ud w
=\left\{
\begin{array}{ll}
\nu_1,&
\textrm{if $W\le \wbh$,}\\
\nu_2,&
\textrm{if $W>\wbh$,}
\end{array}\right.
\label{dens}
\eeq
where 
\beq
\nu_1(W)\equiv \e^{W}\erf\left(\surd W\right)
-\frac{2}{\sqrt{\pi}}\left(W^{1/2}+\frac{2}{3}W^{3/2}
\right)
\eeq
and
\[
\nu_2(W)\equiv 4\pi c\int_{\sqrt{2\wbh}}^{\sqrt{2W}}
\left(W-\frac{w^2}{2}\right)^{1/4}w^2\ud w\, +
\]
\beq
\ \ \nu_1(\wbh)=2^{7/2}\pi c g(W)W^{7/4}+\nu_1(\wbh). \label{nu2}
\eeq
The first term of the r.h.s. of equation (\ref{nu2}) is the density corresponding to a
polytropic stellar model with index $=7/4$ \citep{BT}, with
\beq
g(W)=\int_0^{\,\theta}\left( \sin^{3/2}y-\sin^{7/2}y\right)
\ud y, \label{cw}
\eeq
where $\theta\equiv \cos^{-1}\sqrt{\wbh/W}$, $0\le \theta\le \pi/2$.

The integral in equation~(\ref{cw}) yields
\[
g(W)= -\frac{4}{21}\left[F\left(\frac{\pi}{4}-\frac{\theta}{2}\right)
-F\left(\frac{\pi}{4}\right)\right]-
\]
\[
\ \ \frac{1}{42}(5\cos \theta +3\cos3\theta)\surd \sin\theta=
\]
\beq
\ \ -\frac{4}{21}\left[F\left(\frac{\pi}{4}-\frac{\theta}{2}\right)
-F\left(\frac{\pi}{4}\right)+
(3\omega-1)(1-\omega)^{1/4}\frac{\surd \omega}{2}\right],
\eeq
where $\omega\equiv\wbh/W$ and $F(\phi)\equiv
\int_0^\phi(1-2\sin^2y)^{-1/2}\ud y$
is an elliptic integral of the first kind that
can be accurately evaluated by standard
numerical iterative procedures \citep[see, e.g.,][sect. 6.11]{recipes}.

As in standard King models, it is convenient to impose the relation
\beq
\tilde r^2=\frac{9\sigma^2}{4\pi G\tilde\rho}. \label{rking}
\eeq
With this choice equation~(\ref{poisson0}) becomes
\beq
\frac{d^2 W}{dx^2}+\frac{2}{x}\frac{dW}{dx}=-9\nu(W),
\textrm{ for }x>0,
\label{poisson}
\eeq
that has a family of solutions depending only on the boundary conditions
(with $x_0>0$),
\beq
\begin{array}{lcl}
\frac{dW}{dx}(x_0)&=&W'_0,\\
W(x_0)&=&W_0.
\end{array}
\label{boun}
\eeq
Choosing $x_0=\xbh$, these conditions define a two-parameters $(\wbh', \wbh)$ set
of models\footnote{In standard King models $W'_0\equiv 0$ at $x_0=0$}, with all the
others being just scaling parameters.

By the definition of BHIR it can be assumed that
\beq
\frac{d\Psi}{dr}(\xbh)=\frac{G\M}{(\xbh\tilde r)^2}=\frac{G\mu}{\xbh^2}
\tilde \rho \tilde r, \label{force0}
\eeq
that, using equation~(\ref{rking}), leads to
\beq
\wbh'=-\frac{9\mu}{4\pi \xbh^2}.\label{force}
\eeq
Thus, the whole set of models can be conveniently described by the pair
$(\mu, \wbh)$, while the solution inside the BHIR can be found by integrating
equation (\ref{poisson0}) backwards from $\xbh$ to a given minimum radius,
which we generally set equal to $0.1\xbh$.

Finally, $\xbh$ must be a solution of equation
(\ref{xbh0}) with ${\cal M}(x)$ given self-consistently by the model that
uses $\xbh$ itself as a BHIR radius. This is found iteratively, evaluating
the `new' radius $\xbh^\mathrm{new}$ through equation (\ref{xbh0}) with
${\cal M}(x)$ generated employing the previous value of $\xbh$.
The process starts with an initial guessed $\xbh=G\M/3\sigma^2\tilde r$
and is halted when
the difference between two subsequent values
stays within a given tolerance
\textchanged{
(we chose $|\xbh^\mathrm{new}-\xbh|/\xbh<0.1$).
}

\subsection{Inclusion of a mass spectrum}
The continuum stellar mass spectrum of the real cluster can be represented
by a set of $n$ stellar components, each one having a star mass $m_k$, a total
mass $M_k$ and a velocity parameter $\sigma_k$.
The DF of the more realistic multimass case, $f_\mathrm{mm}$, can be given as
a linear combination of the $f$ of equation (\ref{DF}) \citep[see, e.g.,][]
{dacosta, gunn}:
\beq
f_\mathrm{mm}(E)=\sum_{k=1}^n \alpha_k f_k(E)=
\sum_{k=1}^n \alpha_k f(\beta_kE), \label{fmulti}
\eeq
where $\alpha_k$ serve to reproduce the given mass function and
$\beta_k\equiv\sigma^2/\sigma^2_k$.

A more rigorous treatment would require $f_k$ to explicitly depend on $m_k$
through the exponent of $E$ in the DF inside the
BHIR, as a consequence of mass segregation. This dependence was
quantified by \citet{BW77} and then confirmed (at least partially) by
means of Fokker-Planck \citep{murphy91}, $N$-body \citep{bau04b} and
Monte-Carlo \citep{freitag063} simulations.
Nevertheless, the small gain in `theoretical coherence' might be vanified by the
use of a discrete set of stellar masses instead of the real continuous mass
spectrum. Furthermore, the model would be complicated by the need of
separate numerical quadratures in equation~(\ref{dens}) for each mass class.
For these reasons, we simplified the model construction by considering the same
$1/4$ energy exponent in $f(E<-\wbh)$, regardless of the stellar mass.
However, the energy equipartition, responsible of mass segregation, can be still
reproduced at the border of the BHIR by a suitable choice of $\beta_k$ (see below).

Thus, the multicomponent Poisson equation has the same form of equation
(\ref{poisson}) apart from $\nu$ that is replaced by
the total density
\beq
\nu_\mathrm{tot}(W)\equiv\sum_{k=1}^n
\frac{\alpha_k}{\beta_k^{\,3/2}}\nu(\beta_k W),
\label{poisson_ii}
\eeq
where $\nu$ is given here by equation (\ref{dens}).

Once a component is arbitrarily chosen as
the `reference' one, say the first one, without loss of generality one can
assume $\sigma=\sigma_1$, i.e. $\beta_1=1$. Then, since the set of
$\alpha_k$ are constrained by the given set of $M_k$, the solutions of
the Poisson equation
are determined just by the boundary conditions (\ref{boun})
at $x_0=\xbh$.
Indeed, the set of $\beta_{k>1}$ can be fixed by the requirement
of energy equipartition in the central region. To do that, the procedure described in
\citet{mio} is applied at the
radius\footnote{Because only for $x>\xbh$ all $f_k$ are
lowered Maxwellians.} $\xbh$.
In practice, once given $\sigma$ and $\wbh$, we have
\beq
\beta_k=\wbh^{-1}\kappa^{-1}\left[\frac{m_1}{m_k}\kappa(\wbh)\right],
\eeq
with $\kappa$ the function defined in equation (8) of \citet{mio},
and $\sigma^2_k=\sigma^2/\beta_k$.

Thus, $\wbh$ and $\mu$ determine a 2-parameters family of
models [through equation (\ref{force})] also in the multimass case. Note,
furthermore, that $f_k$ does now depend on $m_k$, through $\beta_k$.

As regards $\xbh$, the same iterative procedure described in the single-component case is
employed to find the value that solves equation (\ref{xbh0}). Finally,
as a consistency test, the velocity dispersion profile generated
by the model is compared with that deduced by the spherical
isotropic Jeans' equation
\citep{BT},
\beq
\media{w^2}_J(x)=\frac{27}{4\pi\nu_\mathrm{tot}(x)}\int_x^{x_\mathrm{t}}\nu_\mathrm{tot}(y)
[{\cal M}(y)+\mu]y^{-2}\ud y,
\eeq
and the model output $[{\cal M}(x),\nu_\mathrm{tot}(x)]$ is accepted only if the
\textchanged{
difference between the two profiles (evaluated as a $\chi^2$ summation over the
spatial grid points) is small enough
(e.g., to within 90 percent of confidence level).
}
\section{Surface brightness profiles with IMBH}
\label{profiles}
\begin{figure}
\includegraphics[width=8.5cm]{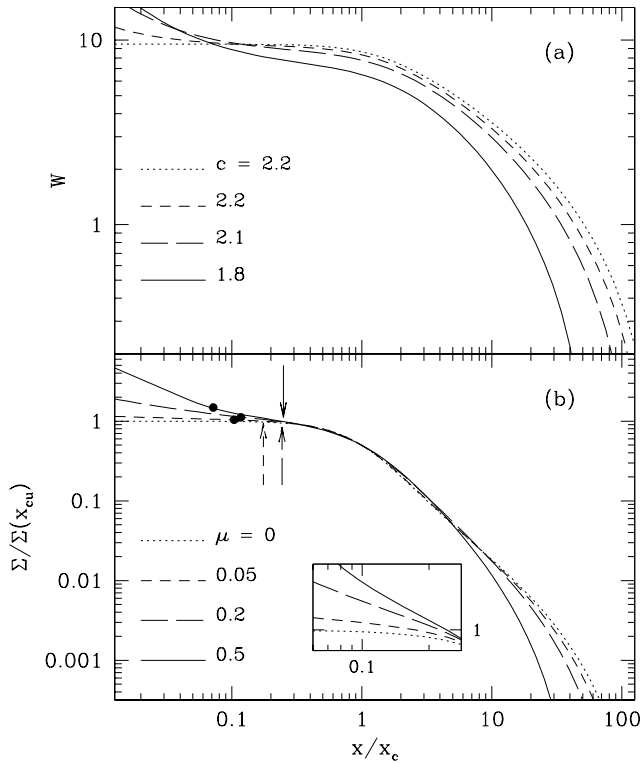}
\caption{(a) Behaviour of the potential and (b) projected surface density profile
(scaled so as to give $\Sigma=1$ at the cusp radius) for single-mass models
with $\wbh=9.5$ and with various IMBH mass as indicated ($\mu =0$ means
a standard King model, with $W_0=9.5$). The dots mark the BHIR radii, while the arrows
indicate the cusps radii ($x_{\mathrm{cu}}$).
The panel (b) inset shows the
profiles in the core region around $x_{\mathrm{cu}}$.
The concentration parameters are also reported in panel (a).
\label{single}}
\end{figure}

\begin{figure}
\includegraphics[width=8.5 truecm]{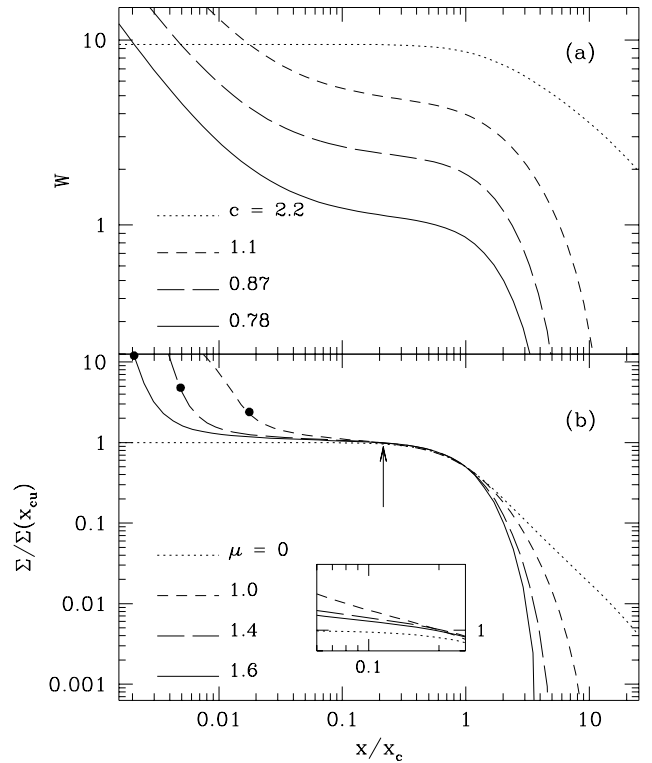}
\caption{The same as in Fig.~\ref{single} for models with higher BH mass.
Since in this case the cusp radii are nearly equal, only one arrow is plotted.
\label{single_hm}}
\end{figure}

To understand how the presence of an IMBH influences
the shape of a typical projected profile, it is instructive to
examine the results of a single-mass model first.
In Fig.s~\ref{single} and \ref{single_hm}, various projected surface density profiles ($\Sigma$)
are plotted,
fixing $\wbh=9.5$, for some of the models written on Table~\ref{tab_single}.
As a comparison, a standard King model is also plotted with $W_0=9.5$.

\begin{table}
 \centering
  \caption{Relevant parameters for single-mass models generated with $\wbh=9.5$. Some
  of them are plotted in Fig.s~\ref{single} and \ref{single_hm}.
  $\mu$ is the dimensionless BH mass, $\M/M$ is the BH mass in units of total cluster mass, $c$
  is the concentration defined according to Equation~(\ref{conc}), $\xbh$ is
the BHIR radius, $x_\mathrm{cu}$ is the cusp radius, $x_\mathrm{c}$ is the core radius
and $s$ is the logarithmic slope of the projected surface density around $x_\mathrm{cu}$.
\textchanged{
In the last line $x_\mathrm{cu}>x_\mathrm{t}$ (i.e. $c<0$) indicates the presence of a
full cuspy profile (in this case $x_\mathrm{c}$ is a meaningless quantity).
}
\label{tab_single}}
  \begin{tabular}{@{}ccccccc@{}}
  \hline
$\mu$ &
$\M/M$&
$c$&
$\xbh$&
$x_\mathrm{cu}$&
$x_\mathrm{c}$&
$s$\\
\hline
no BH&
$0$&
$2.2$&
$0$&
$0$&
$0.92$&
$0$\\
$0.05$&
$5.9\times 10^{-4}$&
$2.2$&
$0.10$&
$0.18$&
$1.0$&
$7.4\times 10^{-2}$\\
$0.2$&
$2.4\times 10^{-3}$&
$2.1$&
$0.15$&
$0.30$&
$1.2$&
$0.18$\\
$0.5$&
$6.0\times 10^{-3}$&
$1.8$&
$0.16$&
$0.56$&
$2.2$&
$0.25$\\
$0.8$&
$8.4\times 10^{-3}$&
$1.4$&
$0.16$&
$1.1$&
$4.6$&
$0.19$\\
$1$&
$8.9\times 10^{-3}$&
$1.1$&
$0.15$&
$1.9$&
$8.5$&
$0.14$\\
$1.4$&
$9.0\times 10^{-3}$&
$0.87$&
$0.14$&
$6.0$&
$28$&
$9.8\times 10^{-2}$\\
$1.5$&
$9.3\times 10^{-3}$&
$0.80$&
$0.14$&
$8.8$&
$42$&
$8.8\times 10^{-2}$\\
$1.6$&
$1.0\times 10^{-2}$&
$0.78$&
$0.14$&
$14$&
$66$&
$7.7\times 10^{-2}$\\
$1.8$&
$1.8\times 10^{-2}$&
$0.77$&
$0.14$&
$84$&
$330$&
$0.12$\\
$2$&
$9.2$&
$<0$&
$0.14$&
$>x_{\mathrm{t}}$&
&
$>1$\\
\hline
\end{tabular}
\end{table}

As expected, we can see that the BH gives rise to a cuspy behaviour
extended from the centre up to a radius that can be called
the `cusp' radius, $x_{\mathrm{cu}}$, above which the much flatter
core region begins.
\textchanged{
An `operative definition' of this radius can be given as the radius where the
surface brightness profile shows a concavity change (while $x_{\mathrm{cu}}\equiv 0$ if
the inflection point is not evident,
i.e. if the cusp is not observed). However, 
}
this radius
was found to be well approximated  -- at least
in the range of BH masses studied here ($\M/M\la 2$ per cent) --
by the radius containing half the BH mass in stars
[${\cal M}(x_{\mathrm{cu}})=\mu/2$].

Moreover, it is convenient to define the `core' radius $x_\mathrm{c}$
as equal to the radius at which the surface brightness drops to $1/2$
the value at $x_{\mathrm{cu}}$.
Indeed, it is important to adopt a concentration parameter that is unambiguously
defined when a density cusp is present.
In particular, the parameter
\beq
c\equiv\log\left(\frac{x_\mathrm{t}}{x_\mathrm{c}}\right) \label{conc}
\eeq
is preferable to the standard King concentration parameter
$c_\mathrm{K}=\log(x_\mathrm{t})$,
because this latter is based on
the length-scale $\tilde r$ that has no immediate relation with any
relevant morphological feature of our IMBH models\footnote{On the contrary,
in King models $x_\mathrm{c}\simeq 1$, i.e. $\tilde r \simeq$ the core radius.}.
\textchanged{
However, it has been checked that an IMBH model with a given $c$ yields
a profile that is well matched, in the core and in the tidal region, by
that of a standard King model with $c_\mathrm{K}=c$.
Note, also, that in the profiles depicted in Fig.s~\ref{single} and \ref{single_hm},
$x_\mathrm{c}$
corresponds to the location of the `knee' of the
curves, in analogy to a standard King profile.
Thus, in clusters whose central luminosity cusp is not resolved,
one can infer about the presence of an IMBH directly using the
standard $c_\mathrm{K}$ parameter coming from observations.
}


From these profiles (plotted as
a function of the radius measured in unit of $x_\mathrm{c}$)
three relevant features can be immediately noticed:
\begin{enumerate}
\item the projected density in
the core region is not as flat as in standard King model,
(see inset in Fig.s~\ref{single}b and \ref{single_hm}b)
\item the presence of a central IMBH reduces significantly the concentration of the
profile in comparison with a BH-free model with $W_0=\wbh$, and
\item the concentration \textit{decreases} for increasing $\mu$.
\end{enumerate}
These features are confirmed (as we will see later) in the case of multimass
models, too.

The last two points can be easily understood from the behaviour of the potential
(Fig.s~\ref{single}a and \ref{single_hm}a) that, at $\xbh$, has a different
slope when compared with the (practically flat) King model.
The potential with the BH decreases below $\wbh$, then becomes flatter
in the core (before dropping in the tidal region) leading the system to
a lower concentration profile in comparison with the BH-free case.
This suggests that the presence of a IMBH induces profiles with a large core.

Note, from Table~\ref{tab_single}, the rapid growth of the cusp radius
for $\mu>1$; for $\mu=2$, it becomes even larger than the limiting
radius making, in fact, $\Sigma(x)$ to ``jump'' to a completely
cuspy behaviour (not shown in the Figures), without any recognizable core.
Since this leads to $\M > M$, such kind of configurations are discarded
throughout our analysis, as unrealistic for globular clusters. Therefore
\textchanged{
-- once the validity of the DF of equation~(\ref{DF}) is accepted -- 
}
reasonable
values of the BH mass at the centre of GCs are compatible \textit{only} with
a core-like profile. The density cusp is confined in the very inner region
of the core.

All this is in remarkable agreement with the results of \citet{bau05}
and \citet{trenti07},
obtained by means of collisional $N$-body simulations including IMBH.
These simulations have shown that the presence of
an IMBH initially enhances the rate of exchange of energy in close encounters
among stars moving in the cusp region; this gives rise to a strong cluster expansion
which, in turn, yields final profiles well fitted by King models, with a relatively
low concentration \citep[see also][]{bau04a,bau04b}.

Finally, the apparent contradiction that can be seen in
Fig.s~\ref{single}--\ref{single_hm} of a decreasing BHIR radius
for increasing BH masses, is actually due to the scaling of the radius with
$x_{\mathrm{c}}$ and to the fact
that the core radius grows quite rapidly while $\xbh$ is nearly constant
(see Table~\ref{tab_single}).
The decrease of the ratio $\xbh/x_\mathrm{c}$
is a consequence of the lower and lower concentration that these
models show for growing BH masses. Another consequence is the unexpected
decrease of the logarithmic slope ($s$, with $\Sigma\sim x^{-s}$) of the
surface density around $x_{\mathrm{cu}}$,
that can
be seen for $\mu>0.5$ (see  Table~\ref{tab_single} and Fig.~\ref{single_hm}b inset).

\subsection{The multimass case}
At this point, one could study in more detail the dependence on the BH mass
of both $c$ and the slope of $\Sigma(x)$ in the core region. Nevertheless,
it is more appropriate to make this analysis directly in the multimass context,
which is (for globular clusters) a much more realistic scenario than that of the
single-mass case, because the mass-segregation could heavily change the effect
of the presence of the BH in the brightness profile \citep{bau05}.

Thus, a stellar mass spectrum is included assuming a Salpeter
mass function (${\ud}N\propto m^{-1.35}\ud \log m$) and considering
$7$ mass bins in the range $0.25$ -- $0.83 \msol$, along with
light and heavy remnants, following the prescriptions of \citet{cote95}.
Nevertheless, their lightest
mass class was not included because it is below the lower mass limit
predicted, under energy equipartition, in \citet{mio}.
See Table~\ref{components} for the list of parameters for the adopted
stellar components.
\begin{table*}
 \centering
 \begin{minipage}{112mm}
  \caption{Values for the components used in the multimass model. MS = main sequence stars; G = giants;
HB = horizontal branch stars; WD = white dwarfs \citep[from][]{cote95}.
  \label{components}}
  \begin{tabular}{@{}ccccccl@{}}
  \hline
$k$ & mass range $(\msol)$               &  $M_k/M$             & $m_k (\msol)$& $(L/M)_k$              & $\sigma_k/\sigma$ & content \\
\hline
$1$ & $0.75$--$0.83$            & $4.9\times 10^{-2}$ & $0.79$       &                $10$& $1$               & MS, G, HB\\
$2$ & $4$--$8$\footnote{Progenitors mass range that yield $1.2 \msol$ WD.}       & $3.9\times 10^{-2}$ &  $1.2$       &                 $0$& $0.81$            & WD\\
$3$ & $0.65$--$0.75$\footnote{Including a $0.7 \msol$ WD population from progenitors with mass $1.5$--$4 \msol$.}  &              $0.18$ & $0.70$       &              $0.19$& $1.1$             & MS + WD\\
$4$ & $0.55$--$0.65$            & $9.3\times 10^{-2}$ & $0.60$       & $6.5\times 10^{-2}$& $1.2$             & MS \\
$5$ & $0.45$--$0.55$\footnote{Including a $0.5 \msol$ WD population from $0.83$--$1.5 \msol$ progenitors.} &              $0.24$ & $0.50$       & $2.3\times 10^{-2}$& $1.3$             & MS + WD\\
$6$ & $0.35$--$0.45$            &              $0.16$ & $0.39$       & $1.0\times 10^{-2}$& $1.6$             & MS\\
$7$ & $0.25$--$0.35$            &              $0.24$ & $0.29$       & $4.9\times 10^{-3}$& $5.7$             & MS\\
\hline
\end{tabular}
\end{minipage}
\end{table*}

\begin{figure}
\includegraphics[width=8.5cm]{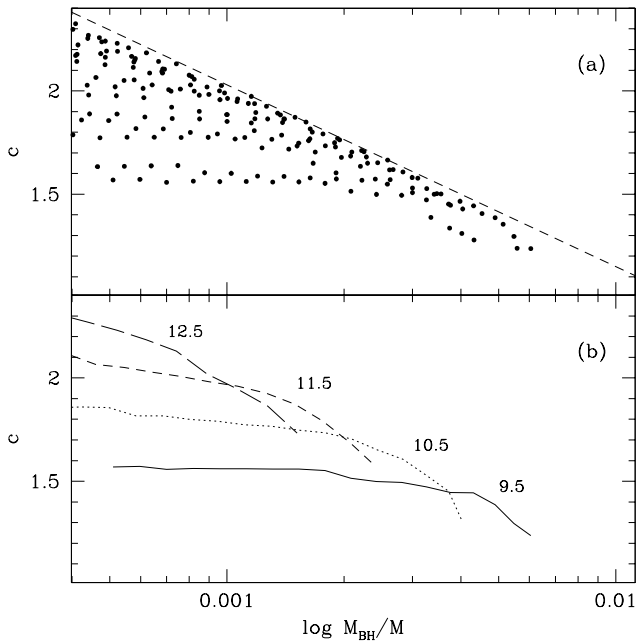}
\caption{(a) Concentration parameter obtained for a grid of values of $\M/M$,
and $\wbh$; no realistic models can be generated above the dashed line. (b)
Behaviour of $c$ as a function of BH mass for some values
of $\wbh$, as indicated.
\label{multi_c}}
\end{figure}
\begin{figure}
\includegraphics[width=8.5cm]{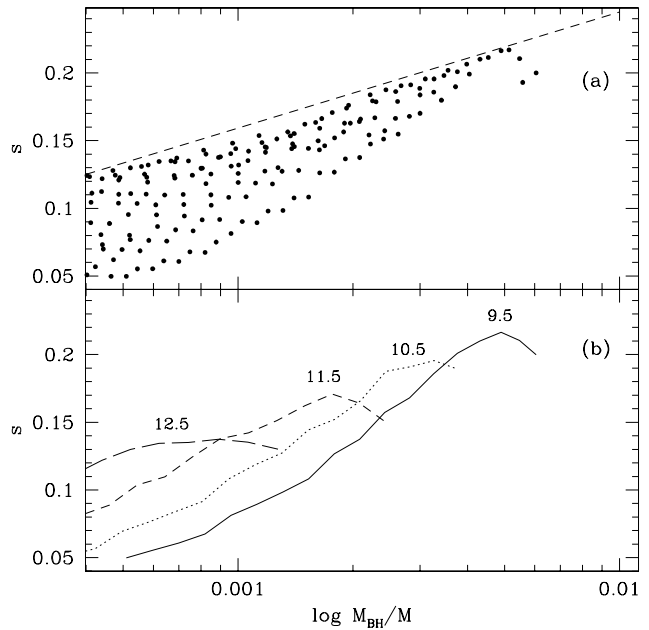}
\caption{As in Fig.~\ref{multi_c} for the logarithmic slope of the surface brightness
at the cusp radius.
\label{multi_smu}}
\end{figure}
The qualitative behaviours of the apparent concentration and of the $\Sigma$ slope
as a function of the BH mass discussed in the single-component case, are basically confirmed by
the curves plotted in Fig.s~\ref{multi_c}b--\ref{multi_smu}b. They refer to a grid of models
computed in the range $15 \leq \wbh \leq 9.5$ and $10^{-2} \leq \mu \leq 0.15$.
In this case the concentration has been evaluated on the total projected surface brightness
profile, $I(x)$, and the logarithmic slope $s$ is again such that $I\sim x^{-s}$.
It is worth noting that:
\begin{enumerate}
\item $c$ decreases monotonically with the BH mass
\item according to our model, IMBHs with $\M\ga 10^{-3}M$ can exist only for
non-collapsed distributions ($c\la 2$) exhibiting a logarithmic slope
in the core which is $\la 0.25$ for reasonably
low BH masses ($\M\la 10^{-2}M$)
\item $s$ increases for growing BH mass up
to a maximum value, then the lower concentration achieved makes the core region
to be much wider than the cusp region, causing the decrease of $s$.
\end{enumerate}
These features fully confirm the \citet{bau05} findings.
\textchanged{
In particular, the logarithmic slope is compatible with the range
of values found for the clusters studied by these authors'
$N$-body simulations. 
}

In Fig.~\ref{multi_c}b it is apparent a certain degree of self-similarity of the various curves;
of course, lower concentrations can be in principle obtained starting from lower $\wbh$
-- though this imposes an even higher mass cut-off of the stellar mass function at the
low-mass end in order to ensure energy equipartition.
Thus, no lower limit on $c$ can be fixed. On the contrary, an upper limit to the
concentration exists (roughly indicated by the dashed line in Fig.~\ref{multi_c}a).
The same argument holds for $s$ (Fig.~\ref{multi_smu}), for which lower values could be
obtained for lower $\wbh$. Therefore, combining the upper limits of the two quantities, a range
of admitted values is found for the BH mass:
\beq
11.6s-4.85\la\log \left(\frac{\M}{M}\right)\la -1.14c - 0.694.
\label{range}
\eeq

\subsubsection{The effect of mass segregation}
\label{masssegregation}
\begin{figure}
\includegraphics[width=8.5cm]{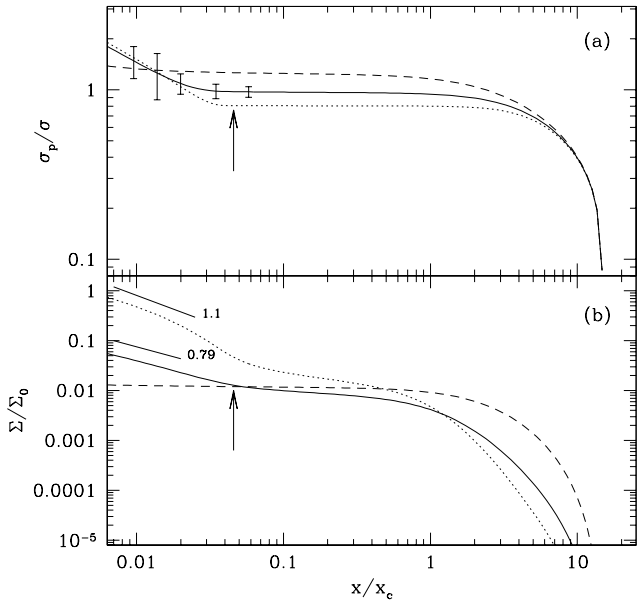}
\caption{(a) Line-of-sight velocity dispersion profile and (b) projected
surface density (normalized to the central value of the heavy remnants
component) for the multimass model described in the text.
Three components are shown: giant and turnoff stars (solid line),
heavy dark remnants (dotted), lightest main sequence stars (dashed).
The central logarithmic slope of $\Sigma$ are also reported for
giants and heavy remnants.
The vertical arrows indicate the BHIR radius.
The error bars refer to simulated velocity dispersion measurements (see text).
\label{mass_segr}}
\end{figure}

The presence of mass segregation (as due to the imposed energy equipartition
at $\xbh$) is evident in Fig.~\ref{mass_segr}. Three components
are represented: the brightest component, the heavy remnants one
and the lightest MS stars component ($k=1,2$ and $7$, respectively,
in Table~\ref{components}).
The first component gives almost
all the cluster luminosity.
The model has been generated with $\wbh=9.5$ and $\mu=0.155$,
corresponding to $\M/M=6\times 10^{-3}$.

In the core region the lightest (heaviest) stars have the highest
(lowest) velocity dispersion and, consequently, the least (most)
concentrated density profile.
The lightest star component is practically unaffected
by the presence of the BH, while inside the BHIR the density cusp
is particularly evident for the most massive stars.

We note that the space density of each component must have the same asymptotic
\citetalias{BW} behaviour $\sim x^{-\gamma}$, with $\gamma=7/4$ (i.e.
$\sim W^{7/4}$),
because it was assumed for simplicity that all the $f_k$ in equation~(\ref{fmulti}) have the
same functional form.
Nevertheless, how much deep one has to go towards the centre to find
the $x^{-7/4}$ cusp depends on $\beta_k$ and, in turn, on the component
stellar mass. This is because the density of the $k$th component is a function
of $\beta_kW\propto \sigma^{-2}_k$  [equation (\ref{poisson_ii})];
thus, the higher the velocity dispersion of the stars is, the deeper the
extent of the region where the `isothermal plateau' prevails, and the flatter the
density is at a fixed radius.

\begin{figure}
\includegraphics[width=8.5cm]{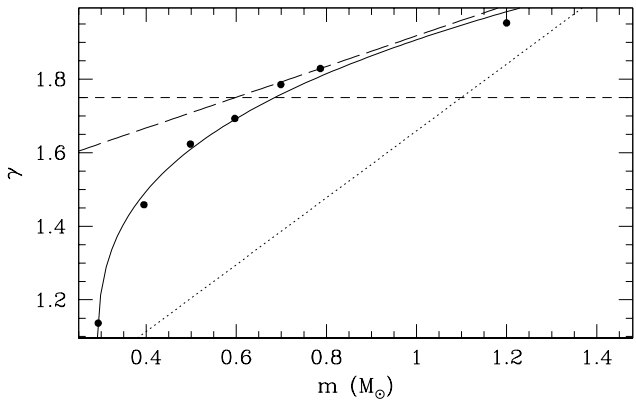}
\caption{Logarithmic slope of the space density of stars evaluated for
the model stellar components (dots) inside the cusp region (at $x= 0.01x_\mathrm{c}$)
and fitted by the law $(m-0.29)^{0.33}+1$ (solid line). The short-dashed
line is the $7/4$ value predicted in the single-mass context. The
fit from the \citet{bau04b} results (dotted line) and the \citet{BW77} relation
(long-dashed) are also reported.
\label{mass_slope}}
\end{figure}

With this in mind, in Fig.~\ref{mass_slope} the log slope $\gamma$ of the
space mass density in the cusp region is plotted as a function of the stellar mass.
The slope is evaluated well inside the BHIR,
namely at $\sim \xbh/5$, i.e. where the enclosed mass in stars is of order $10^{-2}\M$.
It can be noticed that sufficiently massive stars ($m\ga 0.6 \msol$) show a slope close
to the linear relation $\gamma \simeq 1.8m/(0.8 \msol)$, of the kind of that predicted by
the multicomponent Fokker-Planck solutions of \citet{BW77} and then confirmed
by \citet{murphy91}.

Given that low mass stars have a cusp region located deep within the core, they exhibit $\gamma <1.5$,
in accordance to what was found in \citet{bau04a}. These authors also found a particularly low
slope for intermediate mass stars (figure \ref{mass_slope}, dotted line), probably because
in their simulations no particles reached the $x^{-7/4}$
regime \citep[see Fig.~\ref{mass_segr} and the discussion in][]{freitag063}.
On the other hand, \citet{hopman06a}, including energy equipartition in the numerical
integration of multicomponent Fokker-Planck equations, reported power-law density
behaviour with a high slope ($\simeq 2$) for massive objects (stellar BHs around
the Galactic supermassive BH).

For the mass range and mass function considered here, we found that the
$\gamma$ -- $m$ relation is well fitted by the law $\gamma=[m/(1 \msol)-0.29]^{0.33}+1$.
Although we cannot draw really general conclusions, it can be
stated that our prescription to enforce energy equipartition generates results
compatible with the findings of the various studies related to
mass segregation phenomena around a massive BH \citep[see also][]{alexander,merritt}.

To give an indication on
the chance of detecting the IMBH by means of observation of velocity
dispersion profiles\footnote{Because of the assumed isotropy, line-of-sight
velocity and proper motion dispersion are equivalent.}, let us suppose
that the model generated here represents a luminous and nearby cluster (with $c\simeq 1.2$),
located at a distance of
$5$ Kpc, with a central surface brightness $I_0\sim 2\times 10^4$ L$_\odot$ pc$^{-2}$.
If its core radius is a typical $r_c\sim 2$ pc, the BHIR radius is $\rbh=0.094$ pc.
Moreover, let us assume that all the luminosity is given
by the brightest component (that with $k=1$ in Table~\ref{components}),
whose stars would have, at that distance, $V\simeq 16$.
The number of these stars inside the BHIR radius can be estimated by
\beq
N_1\simeq I_0 \pi \rbh^2\left[\left(\frac{L}{M}\right)_1 m_1\right]^{-1},
\label{eqn1}
\eeq
which gives $N_1\simeq 70$,
\textchanged{
thus yelding relatively large Poisson fluctuations that would make
difficult to 
resolve the velocity cusp.
}

This is confirmed by the example reported in Fig.~\ref{mass_segr}a, in which
such measurements are done for
a set of annuli around the cluster centre.
The error bars are evaluated by the usual formulas
(see, e.g., \citealt{jones}; \citealt*{mcnamara})
assuming an rms error on
the single star velocity determination of $\simeq 0.7$ km s$^{-1}$
\citep[similar to, e.g., the typical
accuracy level in the FLAMES-GIRAFFE Very Large Telescope observations by][for stars with the same $V$]
{milone06} and treating the model $\sigma_\mathrm{p}$ profile as if it were the observed
velocity dispersion profile with a typical $\sigma=10$ km s$^{-1}$.
In spite of the relatively high $I_0$ and low cluster distance, it is not
straightforward to identify a cusp
in velocity for such a non-collapsed cluster. Even if in a real observation the error
bars could be smaller because of a larger number of stars actually observable,
the detection of a velocity cusp is certainly more difficult than for
high-concentration clusters \citep[see, e.g.,][]{gerssen02,gerssen04,geb05}.

In general, from the grid of multimass
models generated, we found that $\xbh/x_\mathrm{c}\sim
0.07\pm 0.05$, a ratio that will be used in the following section to give an
estimate about $N_1$ for the candidate clusters.

Finally, note that the radius -- measured in units of $x_c$ -- at which the
velocity dispersion is mostly affected by presence of the BH (where it follows
a keplerian behaviour) is $\sim 0.02$ for the luminous component (Fig.~\ref{mass_segr}a).
This factor is equal to about $3\M/M$, in good agreement with the relation found by \citet{bau05}.

\section{Black hole mass estimated from cluster profile}
\label{massest}
Exploiting equation ~(\ref{range}), we can now make a tentative estimate of the IMBH mass
range in clusters, from their concentration\footnote{
\textchanged{
According to what discussed in section~\ref{profiles}, we can assume that $c$ is equal
to the observed standard concentration parameter.
}
} and from the slope
of the central surface brightness profile.
To this purposes, the recent and detailed analysis made by \citetalias{noyola06}
of previous WFPC2/\textit{HST} observations of 38 galactic globular clusters,
provides the necessary set of data.
Moreover, we also considered the controversial case of the cluster G1 in M31,
for which $s$ is estimated by using the recent ACS/\textit{HST} photometric profile
\textchanged{
published in \citet{ma07}\footnote{
\textchanged{
$s$ is evaluated using the central values of the surface brightness
listed in their table 3 and a second order estimate of the
first derivative.}
}}
 (that gives the most recently updated value for $c$, as well)
and the total mass and luminousity taken by \citet{meylan01}.

From the whole list,
\textchanged{
14
}
clusters were excluded
(marked with `$\times$' in the last column of
Table~\ref{tab_noyo}) because they yield, from equation~(\ref{range}), a lower limit ($\Mmin$)
of the BH mass higher than the upper one ($\Mmax$). All of core-collapsed clusters are in this
subset,
together with those having a steep central brightness profile ($s\ga 0.2$) in spite of being not
very concentrated, like, e.g., NGC 6535 (that has $c=1.3$ and $s\simeq 0.5$).
Of course, the presence of a IMBH cannot be completely ruled out for these clusters.
It can only be stated that such a presence is incompatible with the stars phase-space
distribution of the form given by equations (\ref{DF}) and (\ref{fmulti}).
Moreover, there are 18 clusters (marked with `$\circ$') having very low (or negative) central slope,
for which $\Mmin< 100\msol$; conservatively, they
were considered as clusters without IMBH.
This because the model is based on the assumption that $\M$ is much greater
than any single stellar mass, so that one can assume,  in good approximation, the IMBH to be at rest.
Therefore, as regards these clusters, the model cannot
provide sufficiently stringent
and reliable predictions; accurate individual parametric fits would be necessary to this aim.
Note, however, that the \citetalias{noyola06} uncertainties on
the measurements of $s$ are rather large. This error is not taken into account
in the present analysis.

\begin{figure}
\includegraphics[width=8.5cm]{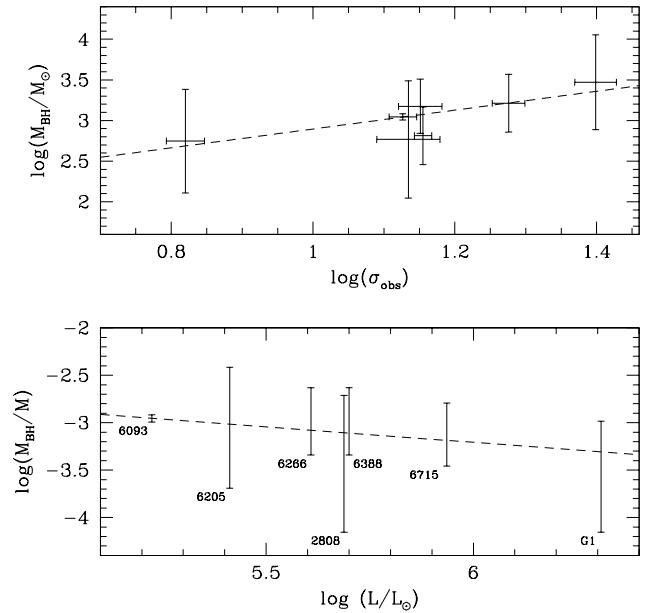}
\caption{Upper panel: BH mass of the candidate clusters
plotted vs. the observed central velocity dispersion (in km s$^{-1}$); lower panel:
log of the ratio $\M/M$ plotted against total cluster luminosity. The dashed lines
are the weighted least-squares fits
\textchanged{
$\log(\M/M)= -1.2-0.33\log(L/$L$_\odot)$ and $\log(\M/\msol)= 1.7+1.2\log(\sigma_\mathrm{obs})$.
}
\label{corr}}
\end{figure}

The relevant parameters of the remaining
\textchanged{
7
}
candidate clusters are written in Table~\ref{tab_noyo2}.
Their predicted IMBH mass range is plotted both vs. the total cluster
$V$-band luminosity and vs. the observed velocity dispersion (Fig.~\ref{corr}).
 It is apparent that $\M/M$ has only a weak dependence
on $L$ while the BH mass tends to increase with $\sigma_\mathrm{obs}$.
To quantify these trends, weighted least-squares fits were computed \citep[following the prescriptions
of][sect. 15.2]{recipes} neglecting, for simplicity, the uncertainties both in $L$ and
in $\sigma_\mathrm{obs}$.
The resulting fits are
\textchanged{
\beq
\log\left(\frac{\M}{M}\right)= (-1.2\pm 1.6)+(-0.33\pm 0.30)\log\left(\frac{L}{\mathrm{L}_\odot}\right) \label{fit1}
\eeq
with $\chi^2_\mathrm{fit}=0.63$ at a confidence level
$P(\chi^2>\chi^2_\mathrm{fit})=0.99$, and
\beq
\log\left(\frac{\M}{\msol}\right)= (1.7\pm 1.4)+(1.2\pm 1.2)
\log\left(\frac{\sigma_\mathrm{obs}}{\textrm{km s}^{-1}}\right) \label{fit2}
\eeq
with $\chi^2_\mathrm{fit}=0.85$ and $P(\chi^2>\chi^2_\mathrm{fit})= 0.97$.
}
The cluster luminosity was chosen as an independent variable because its
measure is much more reliable and directly linked to observations than
the dynamically estimated total masses \citep{pryor93}.

\textchanged{ 
Assuming a uniform global mass-to-light ratio equal to that averaged
among the candidate clusters, $\simeq 3.4(M/L)_\odot$,
equation~(\ref{fit1}) gives $\M\sim 0.09 (M/\msol)^{0.7 \pm 0.3} \msol$.
}
This link is in agreement with the $\M\propto M$ relation between super-massive BHs in galactic nuclei
and the bulge mass \citep{magorrian}, and between the mass of compact
nuclei and that of the host spheroid in nucleated early-type and dwarf galaxies
\citep{cote06,wehner06}.


As regards relation~(\ref{fit2}), it is very different from the law
$\M\sim \sigma_\mathrm{obs}^{4.8}$ found by \citet{ferrarese} for galactic
bulges \citep[see also][]{geb00,tremaine02}.
The reason of this discrepancy can be understood by the same argument
\citeauthor{ferrarese} presented to explain the relation they found:
the $\M$ -- $\sigma_\mathrm{obs}$ link is a consequence of the fundamental
relation $\M\propto M$. In galaxies, this latter relation and the $M\sim L^{5/4}$ and
$L\sim \sigma_\mathrm{obs}^4$ laws lead indeed to $\M\sim \sigma_\mathrm{obs}^5$.
In globular clusters, on the other hand, the observed trends
$M\sim L$ and $L\sim \sigma_\mathrm{obs}^{5/3}$ \citep{heggiemeylan} yield,
through equation~(\ref{fit1}),
\textchanged{
$\M\sim \sigma_\mathrm{obs}^{1.1}$,
}
which is, in fact, compatible with the relation~(\ref{fit2}).

However, since in old globular clusters the secular collisional relaxation has certainly
had a substantial influence on the dynamics within the core region, it is reasonable to expect
only a weak correlation between $\M$ and the presently observed core velocity dispersion.
Moreover, as discussed in
sect.~\ref{masssegregation}, it is rather improbable that $\sigma_\mathrm{obs}$
could provide a really reliable and accurate indication on the velocity dispersion in the inner
cusp region; in fact,
\textchanged{
with the exception of G1,
}
$N_1$ is rather low ($<60$)
for the candidate clusters (see Table~\ref{tab_noyo2}).
Conversely, the cluster total mass, in spite of the tidal erosion, should keep a tighter link
with the environmental conditions at the first stage of the cluster life (at least for
massive clusters), when IMBHs formation are thought to have occurred \citep[e.g.][]{gurkan}.

To improve our capability of identifying the presence of IMBHs, it is important
to search for other indicators that could be related to
the gravitational influence of the compact object on the surrounding cluster stars.
In this context, the presence of an extreme horizontal branch in the
colour-magnitude diagram (CMD) may be a valid indicator, as we will see in the
next Section.

Not surprisingly, however, another indication in favour of the presence of IMBHs
at the centre of our candidate clusters, is that their $r_\mathrm{c}/r_\mathrm{h}$
ratio -- as directly evaluated
from the values listed in the Harris catalogue, see Table~\ref{tab_noyo2} --
is significantly higher than the maximum value ($\sim 0.05$) suggested by \citet{trenti06b}
for clusters that do not harbour IMBHs, and, consequently, do not experience
the enhanced core expansion caused by their presence.

\textchanged{
Note that three over five clusters that \citet{bau05} indicate
as possibly harbouring IMBHs, are in common with our candidates set (namely:
NGC 6093, 6266, 6388). We consider the remaining
two clusters, NGC 5286 and 5694, as non-candidate because they have $\Mmin > \Mmax$
(due to their relatively high concentration and slope); in this respect, notice that the
core slope of the former cluster has been updated to an higher value
in the final \citetalias{noyola06} published list, compared with
that reported by \citeauthor{bau05}. Furthermore, G1, NGC 2808 and 6205 are not
considered by these authors, because they identify as candidate clusters just
those with a non-collapsed core and with a central slope between $0.2$ and $0.3$,
as suggested by the results of their simulations.
}

To conclude this section, it is worth noting the similarity of the surface brightness
profiles reported by \citetalias{noyola06} (plotted in their Fig.~7) with those
illustrated here in Fig.s~\ref{single}b and \ref{mass_segr}b.
By inspecting those corresponding to our candidate clusters, the similarity is rather
evident for M80 and M13. Indeed, these clusters show a clear transition between
a core and a cusp region. Note that they are the only ones, in the \citetalias{noyola06} set,
in which this feature is clearly observed.
Thus, more detailed individual studies, also through parametric fits of the profiles
based on the model discussed here, certainly deserve to be carried out
for them.

\begin{table*}
 \centering
 \begin{minipage}{156mm}
  \caption{Globular clusters in the \citetalias{noyola06} set. $L$ is the total luminosity
in the $V$-band evaluated from $M_V$ listed in the Harris catalogue. $c=$ cluster concentration
(from the Harris catalogue; `:' denotes a core-collapsed cluster). $s$ is the logarithmic
slope \citepalias[from][with opposite sign]{noyola06}. $\Mmin/M$ and $\Mmax/M$ are the
lower and upper limits of $\M/M$ as from equation (\ref{range}).
$M$ is the total cluster mass taken, if available,
from the parametric estimates in \citet{pryor93}, while those marked with a `:'
are estimated by assuming a uniform global mass-to-light ratio $=2$.
EHB$_1$ marked clusters are those with a \citet{harris} HB type $=0$.
EHB$_2$ clusters are those having $\log({T_\mathrm{eff}}_\mathrm{HB})>4.3$
in the \citet{recio06} list.
Last column: $\bullet$ denotes a candidate cluster, `$\times$' a cluster having
$\Mmin > \Mmax$ and `$\circ$' a cluster with $\Mmin < 100 \msol$.
Not available data are indicated by \nod.
\label{tab_noyo}}

\begin{tabular}{@{}lccccccccc@{}}
\hline
NGC number & $\log(L/\mathrm{L}_\odot)_V$ & $c$ &  $s$  & $\log(\Mmin/M)$ & $\log(\Mmax/M)$ &
$\log(M/\mathrm{M}_\odot)$ & EHB$_1$ & EHB$_2$ &case \\
\hline
104   (47 Tuc) & $5.7  $ & $2.03$ & $0    $ & $-4.85 $ & $-3.01$ &  $6.1$ &         &         & $\circ$\\
1851           & $5.26 $ & $2.32$ & $0.38 $ & $-0.442$ & $-3.34$ &  $6.0$ &         &         & $\times$\\
1904   (M79)   & $5.12 $ & $1.72$ & $0.03 $ & $-4.50 $ & $-2.65$ &  $5.2$ & $\surd$ & $\surd$ & $\circ$\\
2298           & $4.45 $ & $1.28$ & $0    $ & $-4.85 $ & $-2.15$ &  $4.8$:&         & \nod    & $\circ$\\
2808           & $5.69 $ & $1.77$ & $0.06 $ & $-4.15 $ & $-2.71$ &  $6.2$ & $\surd$ & $\surd$ & $\bullet$\\
5272   (M3)    & $5.5  $ & $1.84$ & $0.05 $ & $-4.27 $ & $-2.79$ &  $5.8$ &         & \nod    & $\circ$\\
5286           & $5.38 $ & $1.46$ & $0.28 $ & $-1.60 $ & $-2.36$ &  $5.5$ &         & \nod    & $\times$\\
5694           & $5.06 $ & $1.84$ & $0.19 $ & $-2.65 $ & $-2.79$ &  $5.4$ &         &         & $\times$\\
5824           & $5.47 $ & $2.45$ & $0.36 $ & $-0.67 $ & $-3.49$ &  $6.6$ &         & $\surd$ & $\times$\\
5897           & $4.82 $ & $0.79$ & $0.04 $ & $-4.39 $ & $-1.59$ &  $5.1$:&         & \nod    & $\circ$\\
5904   (M5)    & $5.46 $ & $1.83$ & $-0.05$ & $-5.43 $ & $-2.78$ &  $5.6$ &         &         & $\circ$\\
6093   (M80)   & $5.22 $ & $1.95$ & $0.16 $ & $-2.99 $ & $-2.92$ &  $6.0$ & $\surd$ & $\surd$ & $\bullet$\\
6205   (M13)   & $5.41 $ & $1.51$ & $0.1  $ & $-3.69 $ & $-2.41$ &  $5.8$ & $\surd$ & $\surd$ & $\bullet$\\
6254   (M10)   & $4.92 $ & $1.4 $ & $-0.05$ & $-5.43 $ & $-2.29$ &  $5.4$ & $\surd$ & \nod    & $\circ$\\
6266   (M62)   & $5.61 $ & $1.7 $ & $0.13 $ & $-3.34 $ & $-2.63$ &  $5.8$ &         & $\surd$ & $\bullet$\\
6284           & $5.12 $ & $2.5$: & $0.55 $ & $1.53  $ & $-3.54$ &  $5.4$ &         &         & $\times$\\
6287           & $4.88 $ & $1.6 $ & $0    $ & $-4.85 $ & $-2.52$ &  $5.2$:&         &         & $\circ$\\
6293           & $5.04 $ & $2.5$: & $0.67 $ & $2.92  $ & $-3.54$ &  $5.6$ &         & \nod    & $\times$\\
6341   (M92)   & $5.21 $ & $1.81$ & $0.01 $ & $-4.73 $ & $-2.76$ &  $5.3$ &         & \nod    & $\circ$\\
6333   (M9)    & $5.11 $ & $1.15$ & $0    $ & $-4.85 $ & $-2.00$ &  $5.4$:&         & \nod    & $\circ$\\
6352           & $4.52 $ & $1.1 $ & $-0.02$ & $-5.08 $ & $-1.95$ &  $4.8$:&         & \nod    & $\circ$\\
6388           & $5.7  $ & $1.7 $ & $0.13 $ & $-3.34 $ & $-2.63$ &  $6.2$ &         &         & $\bullet$\\
6397           & $4.58 $ & $2.5$: & $0.37 $ & $-0.56 $ & $-3.54$ &  $5.4$ &         &         & $\times$\\
6441           & $5.79 $ & $1.85$ & $0.02 $ & $-4.62 $ & $-2.80$ &  $6.2$ &         &         & $\circ$\\
6535           & $3.83 $ & $1.3 $ & $0.5  $ & $0.95  $ & $-2.18$ &  $4.2$ &         & \nod    & $\times$\\
6528           & $4.56 $ & $2.29$ & $0.1  $ & $-3.69 $ & $-3.30$ &  $4.9$:&         & \nod    & $\circ$\\
6541           & $5.28 $ & $2$:   & $0.41 $ & $-0.094$ & $-2.97$ &  $5.6$ &         & \nod    & $\times$\\
6624           & $4.93 $ & $2.5$: & $0.32 $ & $-1.14 $ & $-3.54$ &  $5.2$ &         &         & $\times$\\
6626   (M28)   & $5.20 $ & $1.67$ & $-0.03$ & $-5.20 $ & $-2.60$ &  $5.4$ &         & \nod    & $\circ$\\
6637   (M69)   & $4.99 $ & $1.39$ & $-0.09$ & $-5.89 $ & $-2.28$ &  $5.3$:&         &         & $\circ$\\
6652           & $4.60 $ & $1.8 $ & $0.57 $ & $1.76  $ & $-2.75$ &  $4.9$:&         &         & $\times$\\
6681           & $4.78 $ & $2.5$: & $0.82 $ & $4.66  $ & $-3.54$ &  $5.2$ &         & $\surd$ & $\times$\\
6712           & $4.93 $ & $0.9 $ & $-0.02$ & $-5.08 $ & $-1.72$ &  $5.0$ &         & \nod    & $\circ$\\
6715   (M54)   & $5.94 $ & $1.84$ & $0.12 $ & $-3.46 $ & $-2.79$ &  $6.3$ &         & \nod    & $\bullet$\\
6752           & $5.02 $ & $2.5$: & $0.03 $ & $-4.50 $ & $-3.54$ &  $5.2$ & $\surd$ & \nod    & $\circ$\\
7078   (M15)   & $5.6  $ & $2.5$: & $0.66 $ & $2.81  $ & $-3.54$ &  $6.3$ & $\surd$ & $\surd$ & $\times$\\
7089   (M2)    & $5.54 $ & $1.8 $ & $-0.05$ & $-5.43 $ & $-2.75$ &  $6.0$ &         & $\surd$ & $\circ$\\
7099  (M30)    & $4.90 $ & $2.5$: & $0.57 $ & $1.76  $ & $-3.54$ &  $5.3$ &         &         & $\times$\\
\textchanged{
G1\footnote{$c$ and $s$ comes from \citet{ma07}; $L$ and $M$ are taken from \citet{meylan01}.}
}
               & $6.31 $ & $2.01$ & $0.06$  & $-4.15 $ & $-2.99$ &  $7.0$ & \nod    & \nod    & $\bullet$\\
\hline
\end{tabular}
\end{minipage}
\end{table*}

\begin{table*}
 \centering
 \begin{minipage}{173mm}
  \caption{Candidate clusters set. $I_0$ is the central surface brightness
estimated from $\mu_V$ in \citet{harris} using equation~(5) in \citet{djorg93}. $r_\mathrm{c}$ is the
core radius in pc, estimated from the angular value and the cluster distance listed in the
Harris catalogue. $\Mmin$ and $\Mmax$ are in $\msol$ and calculated using $M$ from Table~\ref{tab_noyo}.
$\sigma_\mathrm{obs}$ is the central observed velocity dispersion taken from \citet{pryor93}
or, marked with `:', from \citet*{dubath97}. $M/L$ is the mass-to-light ratio in solar
unit. $r_\mathrm{c}/r_\mathrm{h}$ is the ratio
between the core and the half-mass radius, directly estimated from \citet{harris}.
$N_1$ is the estimate of the no. of giants inside $\xbh$
assuming $\rbh=0.07r_\mathrm{c}$ in equation~(\ref{eqn1}).
The other symbols have the same meaning as in Table~\ref{tab_noyo}.
\label{tab_noyo2}}

\begin{tabular}{@{}lcccccccccccc@{}}
\hline

NGC number & $\log(L/\mathrm{L}_\odot)$ & $I_0$ (L$_\odot$ pc$^{-2}$) & $r_\mathrm{c}$ & $\Mmin$ & $\Mmax$ & $M/L$ & EHB$_1$ &
EHB$_2$ & $\sigma_\mathrm{obs}($ km s$^{-1})$& $r_\mathrm{c}/r_\mathrm{h}$ & $N_1$ \\
\hline

2808        & $5.69 $ & $3\times 10^4$    & $0.73$ & $110$             & $3.1\times 10^{3}$ & $3.2$ & $\surd$ & $\surd$ & $13.7 \pm 1.4 $& $0.34$ &$31$\\
6093 (M80)  & $5.22 $ & $2.94\times 10^4$ & $0.44$ & $10^{3}$          & $1.2\times 10^{3}$ & $6$   & $\surd$ & $\surd$ & $13.4 \pm 0.6 $: & $0.23$ &$11$\\
6205 (M13)  & $5.41 $ & $6.68\times 10^3$ & $1.7$  & $130$             & $2.4\times 10^{3}$ & $2.4$ & $\surd$ & $\surd$ & $6.62 \pm 0.41$& $0.52$ &$40$\\
6266 (M62)  & $5.61 $ & $2.54\times 10^4$ & $0.36$ & $290$             & $1.5\times 10^{3}$ & $1.5$ &         & $\surd$ & $14.3 \pm 0.4 $: & $0.15$ &$6$\\
6388        & $5.7  $ & $5.31\times 10^4$ & $0.35$ & $720$             & $3.7\times 10^{3}$ & $3.2$ &         &         & $18.9 \pm 0.8 $& $0.18$ &$13$\\
6715 (M54)  & $5.94 $ & $4.14\times 10^4$ & $0.86$ & $700$             & $3.2\times 10^{3}$ & $2.3$ &         & \nod    & $14.2 \pm 1.0 $& $0.22$ &$59$\\
\textchanged{
G1
}
\footnote{$\mu_V$, $r_\mathrm{c}$ and $r_\mathrm{h}$ comes from \citet{ma07}; $M/L$ and  $\sigma_\mathrm{obs}$ are taken from \citet{meylan01}}
            & $6.31 $ & $1.38\times 10^5$ & $0.78$ & $780$             & $1.1\times 10^4$   & $5.4$ & \nod    & \nod    & $25.1 \pm 1.7$  & $0.12$ &$160$\\ 
 
\hline
\end{tabular}
\end{minipage}
\end{table*}

\section{The IMBH -- EHB connection}
\label{ebtconnection}

Let us assume that the loss of envelope mass is the dominant effect suffered by
a giant during a close passage around the IMBH (`tidal stripping'), and that it
becomes an EHB star because of its uncovered underlying hotter layers
\citep{rich97,alex05}.
Let $r_\mathrm{tid}=R_*(\M/m)^{1/3}$ be the distance from the IMBH
below which the tidal interaction with a star (with radius $R_*$ and mass $m$)
significantly perturbs its internal structure \citep[see, e.g.,][]{freitag02,bau06}.
The order
of magnitude of the rate of tidal stripping events can be
\textchanged{
estimated as
$S\simeq n_0 \sigma_0 \pi r_\mathrm{tid}^2(1+2G\M/r_\mathrm{tid}\sigma_0^2)^2$,
}
where $n_0$ and $\sigma_0$ are, respectively, the typical stellar number density and
velocity dispersion in the cusp region, and where the gravitational focusing is taking
into account in the estimate of the cross-section of the single stripping event
\citep{freitag064}.

For giants having $R_*\sim 100 $ R$_\odot$ and  $m=m_1$, one has $r_\mathrm{tid}\simeq 2.5
\times 10^{-5}$ pc and taking, from the example discussed in section \ref{masssegregation},
$n_0\simeq 3N_1/4\pi\rbh^3\simeq 2\times 10^4$ pc$^{-3}$,
$\sigma_0\simeq 20$ km s$^{-1}$ and $\M=10^3 \msol$, one has
$S\simeq 7\times 10^{-7}$ yr$^{-1}$, which means that of order of
$100$ stripping events can occur within a HB stars
lifetime ($\sim 10^8$ yr), generating a significant population of EHB stars.
Incidentally, this population could have an integrated luminosity ($10^2$ -- $10^4$
L$_\odot$, mainly in the $UV$ band) compatible with the far-UV excess showed by some
-- even metal-rich -- clusters \citep*[e.g. NGC 6388, see][]{rich93},
while, on the contrary, it would be insufficient if one considered
stripping events due to close encounters between \textit{stars} only,
as already pointed out by \citet{rich97}.
Thus, it is reasonable to expect that the existence of an EHB is favoured
by the presence of an IMBH.

To test this hypothesis, we rely on the classification made
in the Harris catalogue (revised on February 2003). This author re-analysed
various set of CMD observations \citep[mainly those by][]{piotto02}, in the intent
of updating the \citet{dickens} HB type, introducing a new type ($=0$) to
account for clusters having a prominent EHB. Thus, clusters with
the HB type $=0$ are considered as having EHB
(they are marked in the EHB$_1$ column of Table~\ref{tab_noyo}).
Then, we study the probability distribution of finding a given number, $N_\mathrm{EHB}$,
of clusters with an EHB, in a set as large as our candidate clusters set.
\textchanged{
However, for the sake of data homogeneity and because of the
lack of direct observations of EHB stars,
to the purpose of this analysis G1 is excluded from the
candidates set.
}

\begin{figure}
\includegraphics[width=8.5cm]{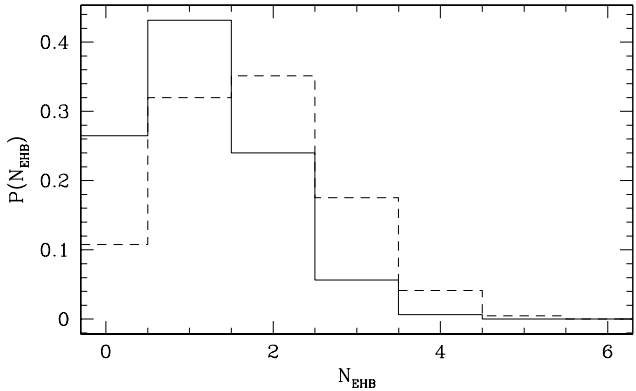}
\caption{Probability distribution of finding $N_\mathrm{EHB}$ clusters having an
extended blue tail, in a subset of 6 clusters randomly
chosen from the \citetalias{noyola06} set. The solid line histogram refers only to
clusters with HB type $=0$, while in the
dashed one those having $\log({T_\mathrm{eff}}_\mathrm{HB})>4.3$
are also taken into account.
Our candidate clusters subset has $N_\mathrm{EHB}=3$ in the
former case and $=4$ in the latter.
\label{p_di_b}}
\end{figure}

A large number of subsets made up of 6 clusters each,
were extracted \textit{at random} from the \citetalias{noyola06} list.
Then, $N_\mathrm{EHB}$ was evaluated for each subset and a frequency distribution
was computed (see the solid histogram in Fig.~\ref{p_di_b}).
The result is that the probability to get \textit{by chance} a set of 6 clusters
with $N_\mathrm{EHB}\ge 3$ (i.e. with $N_\mathrm{EHB}$ equal to or greater than that in
our candidate clusters subset) is $\simeq 0.063$. This, from a statistical
point of view, means that the hypothesis of the \textit{independence} between
the presence of an IMBH and the presence of an extended HB blue tail
can be rejected at a confidence level $\ga 90$ percent.

However, the Harris new HB type was assigned to clusters with an high
relative proportion of blue subdwarfs population in comparison with that at
the redder side of the RR Lyrae region of the HB (Harris, private communication),
whileas the determination of the presence of
the EHB \textit{regardless} of the other HB peculiarities would be more appropriate
to our purposes.
The various parameters defined by \citet{fusi93} to characterize the HB morphology
are unsuitable as well, because their
sample contains only very few clusters of the \citetalias{noyola06} list.
For these reasons, we chose considering the maximum effective temperature,
${T_\mathrm{eff}}_\mathrm{HB}$, evaluated by \citet{recio06} 
for the HB stars of a clusters set that includes a significant part of
the \citetalias{noyola06} sample (see their table~1).

In this case, an EHB was considered as present when $\log({T_\mathrm{eff}}_\mathrm{HB})>4.3$
(i.e. ${T_\mathrm{eff}}_\mathrm{HB}\ga 20,000$ ${}^\circ$K, see e.g. \citealt{moehler};
\citealt*{rosenberg04}).
EHB clusters classified this way are signed in the ETB$_2$ column of Table~\ref{tab_noyo}.
The visual inspection of the \citet{piotto02}
CMDs -- which the \citet{recio06} evaluations are based on -- confirms how in these clusters
the EHB stars are clearly visible and reach downward a $V$ magnitude as faint as the
turnoff point.
While a fraction of the clusters with such an EHB have an Harris HB type $=0$,
further clusters can be considered as owning EHBs.
If one includes these clusters too, our candidates set has $N_\mathrm{EHB}= 4$
and $P(N_\mathrm{EHB}\ge 4)\simeq 0.046$ (see Fig.~\ref{p_di_b},
dashed histogram), consequently the confidence level rises to $\sim 95$ percent.

It is worth mentioning that the two candidate clusters (apart from G1) considered not to
posses an EHB, show significant peculiarities in their HB.
In particular, NGC 6388 shows an over-luminous blue HB \citep{rich97,raimondo02,catelan06}
and, similarly to what observed in M54 \citep{rosenberg04}, a blue HB extending
anomalously beyond the theoretical zero-age HB \citep*{busso04}.
\textchanged{
As regards
G1, we remind that there are some indirect indications suggesting the presence of a
non-negligible population of EBH stars \citep{peterson03} and that,
at least, such a presence cannot be excluded according
to the most recent CMD observations \citep{rich05}.
}

It can be also noticed that in the \citetalias{noyola06} sample, there are 7
non-candidate clusters that possess EHB. This suggests that there might be other
mechanisms -- not related to the interaction with IMBHs -- at work for the formation
of EHB stars. However, 4 of those clusters are highly concentrated ($c>2.4$) or
core-collapsed. This confirms that, whatever the formation mechanism is, it
is favoured by an high density environment \citep{fusi93,buonanno97}.
Nevertheless, the presence of such an environment at the present time is not a sufficient
condition for the formation of an IMBH, which is closely linked to the
\textit{initial} central density instead.
Gravothermal oscillations
and collapse may have changed rapidly the former dynamical state,
thus erasing the link with the initial environment. All this could explain
the rather weak correlation found by \citet{recio06} between
${T_\mathrm{eff}}_\mathrm{HB}$ and the collisional parameter they defined
on the basis of the present dynamical conditions. Furthermore, the velocity
dispersion used in the definition could be not representative of the
higher actual value in the inner (not resolved) cusp region.

However, there are other alternative explanations for the
presence of blue subdwarfs, which are not based upon tidal stripping
\citep[see ][sections 7.3 -- 7.4 and references therein]{catelan05}.
\textchanged{
In this respect, notice that the helium self-enrichment -- probably caused
by multiple episodes of star formation -- that in
some clusters can explain the existence of EHB populations
\citep[e.g., in NGC 2808, see][]{dantona05,piotto07},
could be also related to the presence of an IMBH.
In fact, the IMBH could originate subsequent bursts of star formation, in a way
similar -- though on smaller scales -- to what should occur into the
accretion disk of super-massive BHs \citep[e.g.][]{shlosman89, collin99}.
}

\section{Conclusions}
\label{concl}
In this paper, the multimass isotropic and spherical King model has been
extended to include the \citet{BW} distribution function in the central region,
so as to account for the presence of a central intermediate-mass black hole (IMBH).
This self-consistent model generates surface brightness profiles that show a power-law
behaviour around the centre and a core-like profile outwards.
This latter is similar to a King profile with concentration $c<2$, though slightly
steeper in the core region.
The following features are particularly relevant:
\begin{description}
\item $c$ decreases monotonically with the IMBH mass $\M$;
\item IMBHs with $10^{-3}M\la\M\la 10^{-2}M$, with $M$ the cluster mass,
are compatible only with non-collapsed distributions ($c\la 2$);
\item the core surface brightness profiles exhibit a logarithmic slope
$s\la 0.25$ for reasonably low IMBH masses ($\M\la 10^{-2}M$).
\end{description}

These features are in remarkable agreement with the results of recent collisional
$N$-body simulations, which found that the IMBH causes a quick expansion
of the a cluster core because of the enhanced collisional rate of
exchange of kinetic energy both between stars of different masses \citep{bau05}
and between single stars and hard binaries in 3-body encounters occurring
in the black hole vicinity \citep{trenti07}.
\textchanged{
In particular, the slope of the profile of the core region in the simulated
clusters attains values compatible with those found by our model.
Moreover, the radius inside
which the velocity dispersion profile is appreciably altered by the
influence of the compact object ($\sim 3\M/M$ in units
of the core radius) coincides with the value found by \citeauthor{bau05}.
We can also confirm that a full cuspy density profile (similar to, e.g.,
that of collapsed clusters) can be associated only to a black hole with a
unrealistically high mass, even higher than that of the host cluster itself.
}

A grid of models is then generated to derive possible
trends of the IMBH mass with the morphological parameters.
It is found that, in general, $12s-4.8\la\log (\M/M)\la-1.1c - 0.69$.
This mass range estimate is applied to a set of 38 galactic globular
clusters recently re-analysed by \citet{noyola06}
as well as to the case of G1 (in M31).
It was found that
\textchanged{
seven
}
clusters in this sample probably host an IMBH, namely:
NGC 2808, NGC 6388, M80, M13, M62, M54 and
\textchanged{
G1.
}
Among these, the M80 and M13 brightness profiles presented in \citeauthor{noyola06}
show the typical appearance of the cuspy-core behaviour reproduced by the model.
\textchanged{
It is worth noticing, moreover, that among the five clusters that \citet{bau05}
indicate as possibly harbouring IMBHs, M80, M62 and NGC 6388 are in common with our
candidates set.
}

The scaling relations
\textchanged{
\beq
\M\sim 0.09\left(\frac{M}{\msol}\right)^{0.7} \msol, \label{m-mbh}
\eeq
\beq
\M\sim 50\left(\frac{\sigma_\mathrm{obs}}{\textrm{km s}^{-1}}\right)^{1.2} \msol,
\label{v-mbh}
\eeq
}
are satisfied by our candidate clusters.
Relation (\ref{m-mbh}) is close to the $\M\sim M$ law found in larger scale systems
\citep{magorrian,cote06,wehner06}.
None the less, relation (\ref{v-mbh}) is significantly different for the analogous relation
found in galaxies \citep{ferrarese,geb00,tremaine02}, but the discrepancy can be understood
by noting that the scaling law between the total mass and the
central velocity dispersion for galaxies is very different from that followed by
globular clusters.

Unfortunately, the uncertainties on the mass range estimates are in many cases
rather large and, as a consequence, the presence of IMBHs on other clusters cannot be
completely ruled out. They are mainly due to the uncertainties in the central observed
projected velocity dispersion, and in the logarithmic slope measurements.
For this reason, we plan to carry out individual studies on a representative cluster
subset in the \citet{noyola06} sample, to infer the IMBH mass directly from the
best parametric fit of their brightness profiles.

Nevertheless, this preliminary and `collective' approach, permits to
achieve an intriguing result: the presence of an extreme blue horizontal branch --
determined according both to the \citet{harris} HB type index and to the maximum
effective temperature of HB stars --  is associated to the presence
of the IMBH at a statistically significant level of confidence ($>90$ percent).
Such a correlation is not surprising when one takes into account the high stellar
density in the inner cusp region and the relatively high ratio between $\M$ and
the single stellar mass, in calculating the rate of tidal stripping event in
giants-IMBH close encounters.

This firmly suggests that the mass of the IMBH could be one of the `hidden' parameters
that are being searched for to explain the strong variability of HB morphology
in clusters with the same metallicity and similar CMDs
(the so-called `2nd parameter' problem).
For instance, the presence of a central IMBH
\textchanged{
in M13 and NGC 6388
could explain the origin of their extreme HB stars, while their absence
in the respective `counterparts' M3 and 47 Tuc is consistent with the
the fact that, according to our analysis,
}
IMBHs should not reside in these latter.

\section*{Acknowledgments}
The author would like to thank E. Brocato, G. Raimondo, A. Sills, G. Piotto, F.
D'Antona and W.E. Harris for their helpful suggestions and discussion about
extreme HB stars
and morphology indicators. The author's acknowledgments go also to M. Freitag
for his valuable
comments, and to the referee (E. Noyola) for pointing out new data
available on G1 and for a careful reading of the paper. Finally, the author
is grateful to all the staff at the Osservatorio Astronomico di Teramo
for the kind hospitality and the friendly atmosphere enjoyed during his stay.

\onecolumn

\bsp

\label{lastpage}

\end{document}